\documentclass[11pt,a4paper]{article}
\usepackage[utf8]{inputenc}
\usepackage[T1]{fontenc}
\usepackage[british]{babel}
\usepackage{amsmath, amsfonts, amssymb, graphicx, float, lipsum, xcolor, hyperref}
\hypersetup{hidelinks}
\definecolor{green(ryb)}{rgb}{0.4, 0.69, 0.2}
\usepackage[left=2.00cm, right=2.00cm, top=2.00cm, bottom=2.00cm]{geometry}
\graphicspath{ {./figures/} }

\newtheorem{theorem}{Theorem}[section]
\newtheorem{rmk}{Remark}[section]
\newtheorem{example}{Example}[section]

\title{Verifiable encodings in multigroup fully homomorphic encryption}
\author{Rams\`es Fern\`andez-Val\`encia 
        \\ IOV Labs
        \\ \href{mailto:ramses@iovlabs.org}{ramses@iovlabs.org}}
\date{}

\begin{document}
\maketitle


\begin{abstract}
This article presents the application of homomorphic authenticators, replication encodings to be precise, to multigroup fully homomorphic encryption schemes. Following the works of Gennaro and Wichs on homomorphic authenticators in combination with the work of multigroup schemes by Kwak et al. we present a verifiable solution for a fully homomorphic primitive that includes the multikey, multiparty and single-user cases. Furthermore, we propose a line of prospective research in constrained-resource scenarios.
\end{abstract}


\section{Introduction}
Homomorphic Encryption (HE) allows computations on encrypted messages without decryption. While a fully HE scheme supporting arbitrary computations was an open problem, Gentry's breakthrough \cite{Gentry2009} led to significant progress, including the development of BFV \cite{Brakerski2012, Fan2012} and CKKS \cite{Cheon2017}. HE is suitable for cloud-based environments, as it supports secure computation without the data owner's presence.

However, standard HE has limitations in situations involving more than one user, such as the multiparty setting. In the case of multiple data sources, using a single-key HE leads to an authority concentration issue, as one party gains access to all data. Multiparty HE (MPHE) \cite{Asharov2012} and multikey HE (MKHE) \cite{Lopez2012} are examples of schemes that overcome this limitation by distributing the decryption authority among multiple parties, thus protecting the privacy of data owners.

This paper focuses on a recent primitive called multigroup HE (MGHE) introduced by \cite{KLSW2021}, which offers interaction and flexibility advantages. MGHE is a generalized variant of HE for multiple parties that combines the best of MPHE and MKHE. An MGHE scheme enables a group of parties to collaboratively generate a public key that can be commonly used among the parties for encryption, thus behaving like an MPHE scheme within a single group. Additionally, MGHE supports arbitrary computations over encrypted data, regardless of whether the input ciphertexts are encrypted under the same group key or not, making it similar to MKHE.

Homomorphic multiplication, which is the most expensive operation in HE regardless of the type of scheme, consists of two steps: tensor product and relinearization. Given encryptions $ct = (c_i)_{0 \leq i \leq n}$ and $ct' = (c'_i)_{0 \leq j \leq n}$ of two messages $m$ and $m'$, respectively, it first computes the product $ct'' = (c_{i,j})_{0 \leq i,j \leq n} = (c_i \cdot c'_j)_{0 \leq i,j \leq n}$, which can be viewed as a valid encryption of $m\cdot m'$ decryptable by $s_i \cdot s_j$; then the relinearization procedure follows to convert $(c_{i,j})_{0 \leq i,j \leq n}$ back to the standard form. The total complexity of relinearization grows quadratically with $n$ since the process should be repeated on $c_{i,j}$ for all $0 \leq i, j \leq n$.

In this paper, we design an improved version of the multigroup BFV scheme in \cite{KLSW2021}. It is built upon the works done by \cite{KKLSS2022} and uses the concept of homomorphic gadget decomposition. A gadget decomposition is said to be homomorphic if it supports the computation over decomposed vectors. This means that arithmetic operations can be performed over the gadget decompositions $h(a)$ and $h(b)$ of any elements $a$ and $b$, so that $h(a) + h(b)$ and $h(a) \odot h(b)$ satisfy $\langle h(a) + h(b), g\rangle  = a + b \mod q)$ and $\langle h(a) \odot h(b), g\rangle = a \cdot b \mod q$ where $\odot$ denotes the component-wise product of vectors. Hence, $h(a) + h(b)$ and $h(a) \odot h(b)$ can be considered as valid decompositions of $a + b$ and $a \cdot b$, respectively.

Similar to \cite{KKLSS2022}, our MGHE construction utilizes homomorphic gadget decomposition to avoid performing $n^2$ gadget decompositions for all pairs $(i, j)$. Instead, we separately compute $h(c_i)$ and $h(c'_j)$ for $0 \leq i, j \leq n$ and combine them to form a valid decomposition of $c_i \cdot c'_j$. Furthermore, we avoid the traditional multiplication strategy that involves tensor product and relinearization. Instead, we combine the two steps and redesign the entire multiplication algorithm, enabling each ciphertext to be pre-processed before being multiplied with another ciphertext.

One major drawback of implementing HE schemes in cloud-based environments is the absence of any guarantees for clients regarding the accuracy of computations carried out by the cloud. Although there are typically service-level agreements in place between clients and clouds, some clients may require more reliable assurances and error-detection capabilities to guard against untrustworthy cloud providers who could introduce errors in their sensitive data and applications with ease. For instance, in machine learning setups assisted by cloud computing, a malicious cloud provider may easily introduce a backdoor while training a model for a security-sensitive application, such as malware detection, or return an incorrect prediction that could lead to a misdiagnosis in a medical application.

Although verifiable computation techniques, including those proposed by Bunz \cite{Bunz2018}, Gennaro \cite{Gennaro2010, Gennaro2013}, and others, can be used to verify the integrity of outsourced computations, integrating them with the complex structure of HE schemes remains a challenge. This is particularly true for popular HE schemes such as BFV, which operate over a large polynomial structure that is often incompatible with verifiable computation techniques.

Only a limited number of studies have explored the verification of computations performed on homomorphically encrypted data. Fiore et al. \cite{Fiore2014} developed a solution utilizing homomorphic message authentication codes to verify computations on ciphertexts, focusing on quadratic functions over a particular variant of the BV scheme \cite{Brakerski2011}. Other subsequent works involve compressing the ciphertexts and generating proofs of correct computation using succinct non-interactive arguments of knowledge (SNARKs) \cite{Fiore2020, Bois2020}. Recently, Ganesh et al. \cite{Ganesh2021} proposed a new ring-based SNARK that is more compatible with lattice-based HE schemes.

The previous studies were limited to specific HE schemes and lacked essential functionalities such as relinearization, rotations, and bootstrapping, rendering them theoretical and impractical for many applications. To address these limitations, we propose a solution for verifying outsourced computations based on the works of \cite{Chatel2022, Gennaro2013}. Unlike the SNARK-based approaches that prioritize efficient verification and succinctness, our primary goal is to provide versatility for homomorphic computations with reasonable costs for both clients and the cloud.

As a final contribution we adapt the techniques developed in \cite{Luo2022} to generate a version of our verifiable MGHE without requiring a common reference string. The common reference string (CRS) is used to connect the ciphertexts of all parties under different keys to enable accurate computation. However, in an attempt to eliminate the need for a CRS, Kim, Lee, and Park proposed an alternative scheme in \cite{KLP2018}. In their approach, the parties share their public keys after generating them, establishing relationships among their keys similar to the threshold fully homomorphic encryption scheme \cite{Asharov2012}. It is worth noting that each user generates their keys independently in this scheme.


\section{Basic concepts}

\subsection{Taxonomy of homomorphic schemes}
It is well-known that homomorphic encryption schemes can be classified in terms of the operations admitted. With this point of view, one talks about somewhat, levelled or fully homomorphic schemes. Within each of these categories one can also classify schemes in terms of users, involved keys and how the former and the latter relate. In this direction, one can talk about:

\begin{itemize}
    \item Single-user schemes: these are the most familiar schemes, where each user holds its public, secret and evaluation keys and uses them to communicate with a server in charge of performing computations.
    \begin{center}
    \includegraphics[scale=.25]{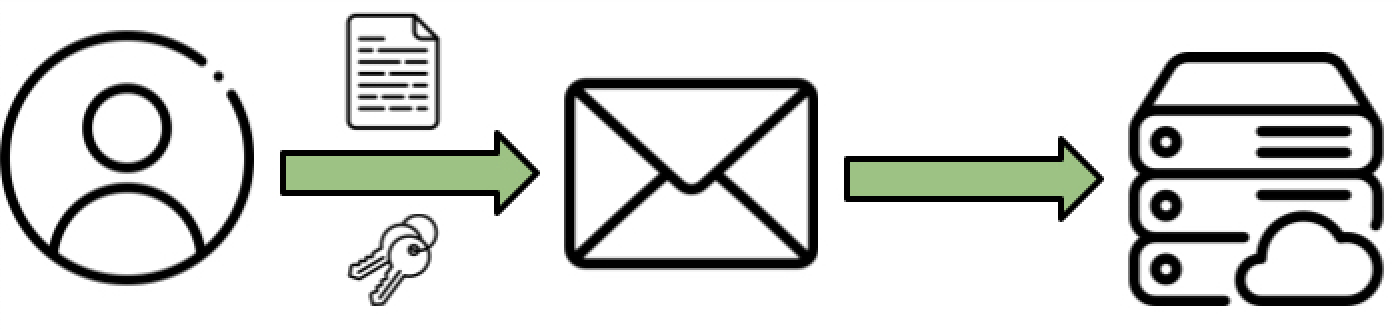}        
    \end{center}

    \item Multiparty schemes: in a multiparty scheme one finds a group of individuals collaborating to the generation of a common public key from their individual keys. In this case, each user holds their own secret key.
    \begin{center}
    \includegraphics[scale=.25]{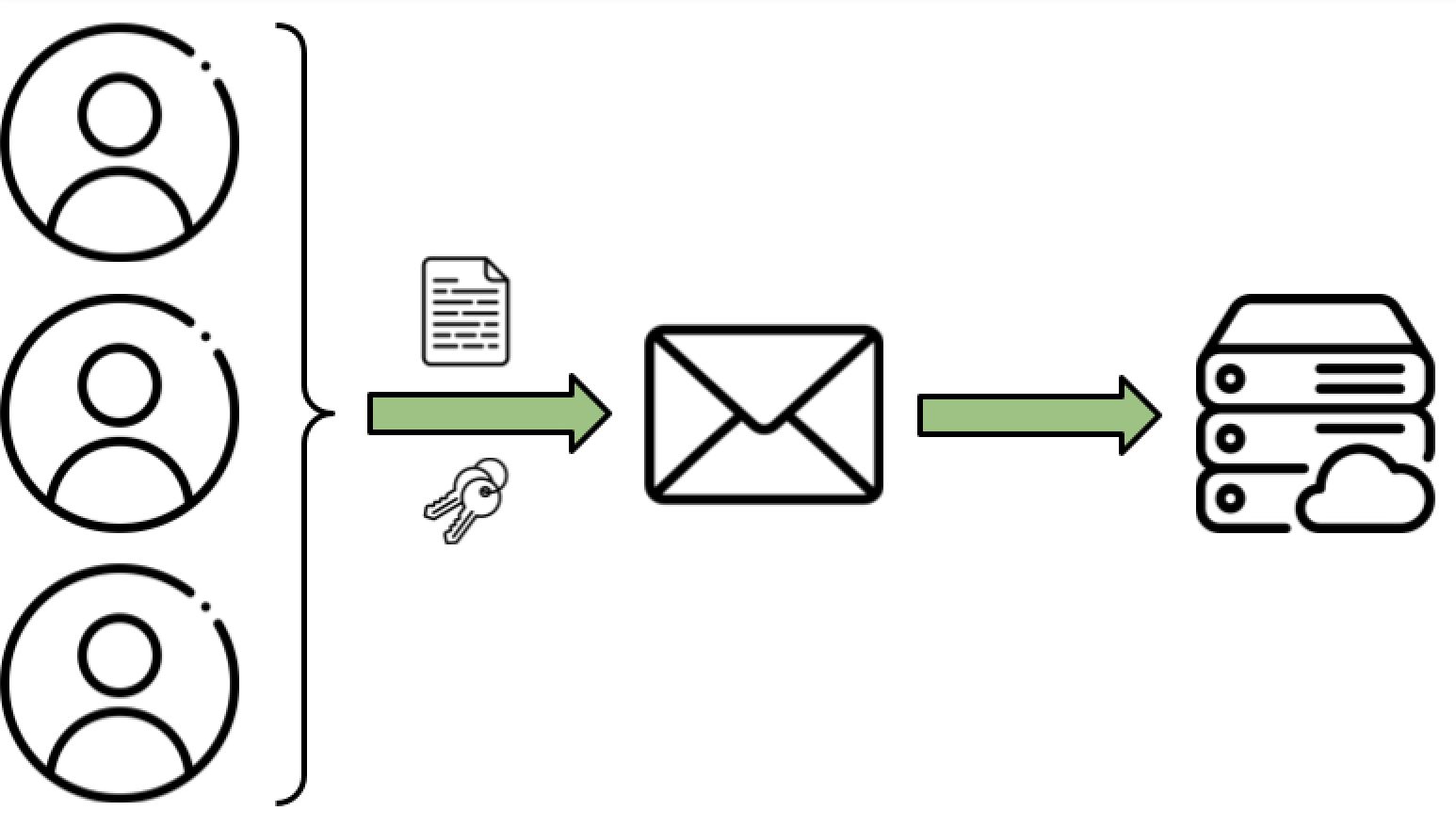}        
    \end{center}
    
    \item Multikey schemes: in a multikey scheme one finds a group of users which send encrypted messages, to be computed by a server, using their own public key. As in the previous kinds of schemes, each user also holds their own private key.
    \begin{center}
    \includegraphics[scale=.25]{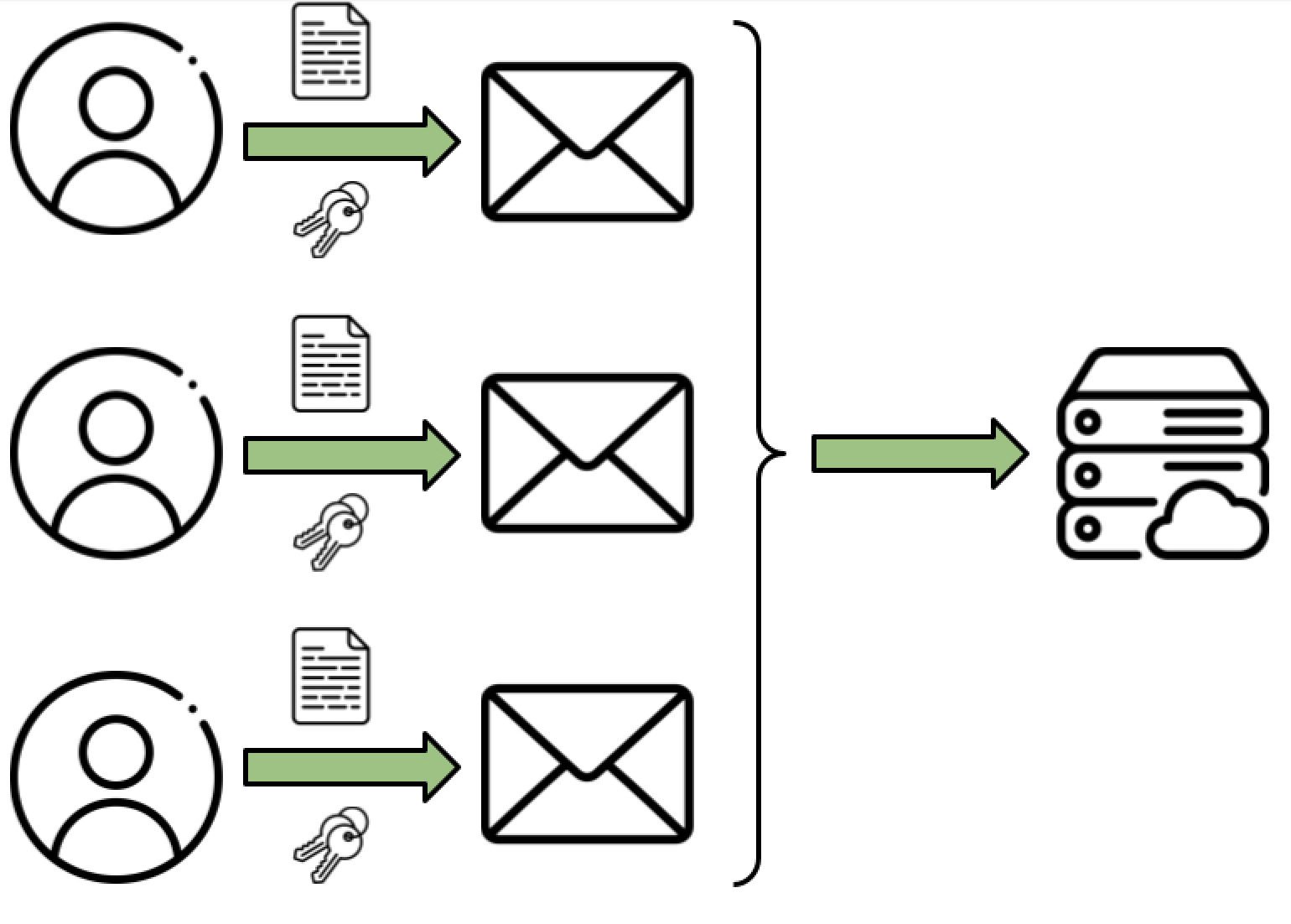}        
    \end{center}
    
    \item Multigroup schemes: this is the most general scenario, where we find different groups of users encrypting messages using a group public key generated in each group using each public key.
    \begin{center}
    \includegraphics[scale=.25]{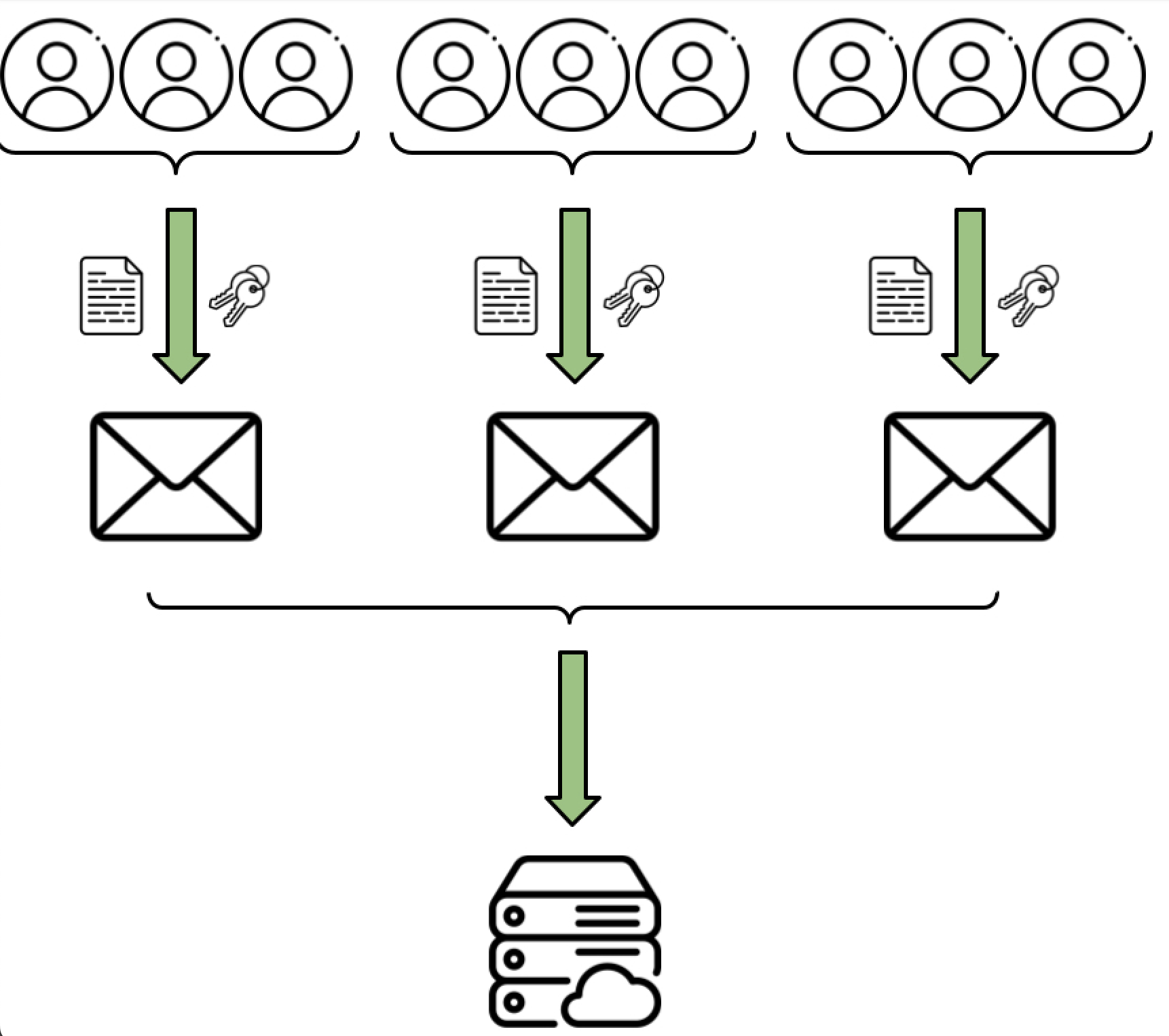}        
    \end{center}
\end{itemize}

\subsection{Labelled programs}
One follows the work done by Genaro and Wichs \cite{Gennaro2013} to formally introduce a labelled program as a tuple $P = (f ; \tau_1, \dots , \tau_n)$ which consists of a circuit $f$ along with distinct input labels $\tau_i \in \{0, 1\}^\star$ for each input wire $i \in \{1, \dots, n\}$. 

 We define the composed program as $P^\star = g(P_1, \dots, P_t)$, given some labelled programs $P_1, \dots, P_t$ and a circuit $g$. It corresponds to evaluating $g$ on the outputs of $P_1, \dots, P_t$. The labelled inputs of $P^\star$ are all the distinct labelled inputs of $P_1, \dots P_t$, meaning that one collects all the input wires with the same label and converts them into a single input wire.

The identity program with label $\tau$ is $\text{Id}(\tau) = (g_{\text{Id}}, \tau)$, where $g_{\text{Id}}$ is the canonical identity circuit for some label $\tau$. Any program $P = (f ; \tau_1, \dots , \tau_n)$ can be written as the composition of identity programs $P = f(\text{Id}(\tau_1), \dots, \text{Id}(\tau_n))$.

\subsection{Hash trees associated with circuits}
For the generation of the authenticator, it will be required to use a concept similar to Merkle trees. Let us consider a function $f$, modelled as a circuit, together with a hash function $H$. The hash tree of $f$ is a Merkle tree sharing the circuit structure of $f$ but replacing the internal gates of $f$ with the hash function $H$. To be precise, let $\mathcal{M}$ be a plaintext space, let $H: \mathcal{M} \to \mathcal{M}$ be a hash function and $f: \mathcal{M}^n \to \mathcal{M}$ a function modelled by a circuit, the has tree $f^H: \mathcal{M}^n \to \mathcal{M}$ is a function taking messages $m \in \mathcal{M}$ for each input wire of $f$. For every wire $w$ in the circuit $f$, the value of $f^H(m_1, \dots, m_n)$ at $w$ is:
\begin{enumerate}
    \item $val(w_i) = H(m_i)$ if $w_i$ is the i-th input wire of $f$.
    \item $val(w) = H(val(w_1), \dots, val(w_t))$ if $w$ is the output wire of some gae with input wires $w_1, \dots, w_t$.
\end{enumerate}

We define the output of the function $f^H(m_1, \dots, m_n)$ as the value of the output wire of $f$.

\subsection{Gadget decompositions}
Let us consider, for an integer $N > 0$, the ring of polynomials $R = \mathbb{Z}[x]/(x^N + 1)$, for $q\in\mathbb{Z}$ one writes $R_q = \mathbb{Z}_q[x]/(x^N + 1)$. A gadget decomposition, with gadget vector $g\in\mathbb{Z}^k$, is a function $h: R_q \to R^k$ which transforms an element $ a\in R_q$ into a $k$-dimensional vector $u = (u_0, \dots, u_{k-1})\in R^k$ of small polynomials such that $a = \langle g, u \rangle \mod q$ for $\langle\,,\,\rangle$ the inner product.

\begin{example}
A classical example of gadget decomposition is binary decomposition, where the gadget vector is given by $g = (1, 2, 2^2, 2^3, …, 2^k)$ for some $k > 0$.
\end{example}

\begin{example}
The ciphertext modulus $Q$ can be chosen as a product $Q = \prod_{0 \leq j < l} q_j$ of pairwise coprime integers $q_0, \dots, q_l$. The Chinese remainder theorem then states the isomorphism $R_Q \xrightarrow{\cong} \prod_{0 \leq j < l} R_{q_j}$ such that $a \mapsto ([a]_{q_j})_{0 \leq j < l}$. We call $([a]_{q_j})_{0 \leq j < l}$ the RNS representation of $a \in R_Q$ with respect to the base $\{q_0, \dots, q_l\}$.  
\end{example}

A gadget decomposition $h: R_q \to R^k$ is homomorphic if the following holds, for $a, b\in R_q$:
\begin{align} 
\langle h(a) + h(b), g\rangle = a + b \mod q \\ 
\langle h(a) \odot h(b), g\rangle = a \cdot b \mod q
\end{align}

where $\odot$ represents the component-wise product of vectors.

Let us introduce the following RNS-friendly gadget decomposition: for $\{q_0, \dots, q_{l-1}\}$ a set of primes, $0 = j_0 < j_1 < \dots < j_k = l$, and $0 \leq i < k$ we introduce the partial products 
\begin{equation}
    Q_i = \prod_{j_i \leq j < j_{i+1}} q_j
\end{equation}

These products are pair-wise coprime integers satisfying
\begin{equation}
    q = \prod_{0 \leq i < k} Q_i    
\end{equation}
This leads to the introduction of a gadget vector $g = (g_0, \dots, g_{k-1})\in R_q^k$ such that, if $[\star]_q$ denotes the reduction modulo $q$:
\begin{equation}
    g_i = \left[ \left( \prod_{i' \neq i} Q_{i'} \right)^{-1} \right]_{Q_i} \cdot \left( \prod_{i' \neq i} Q_{i'} \right)    
\end{equation}
One observes that this vector $g$ satisfies that $g_i = 1 \mod q_j$ for $j_i \leq j < j_{i+1}$, and $g_i = 0 \mod q_j$ otherwise. 

One defines the digit decomposition as $h: R_q \to R^k$ such that
\begin{equation}
h(a) = \left( [a]_{Q_0}, \dots, [a]_{Q_{k-1}}\right)    
\end{equation}
The digit decomposition is a gadget decomposition associated with $g$, and it has the homomorphic properties introduced previously. It is important to note that in the situation where $k = l$ and $Q_i = q_i$ for $0 \leq i <l$, $h$ is called the prime decomposition.

Given $0 \leq i < k$, for any $a \in R_q$ in the RNS form $a_j = [a]_{q_j}$, each component $[a]_{Q_i}$ of the digit decomposition can be written as
\begin{equation}
   [a]_{Q_i} = \sum_{j_i \leq j < j_{i+1}} \left[ \frac{q_j}{Q_i} \cdot a_j \right]_{q_j} \cdot \frac{Q_i}{q_j} - Q_i \cdot \left\lfloor \sum_{j_i \leq j < j_{i+1}} \left[ \frac{q_j}{Q_i} \cdot a_j \right]_{q_j} \cdot q_j^{-1} \right\rceil 
\end{equation}
The above expression holds in $R$, so it can be used to compute the RNS form of $[a]_{Q_i}$ over an arbitrary modulus $p$, indeed:
\begin{equation}
    \left[ [a]_{Q_i} \right]_p = \sum_{j_i \leq j < j_{i+1}} \left[ \frac{q_j}{Q_i} \cdot a_j \right]_{q_j} \cdot \left[\frac{Q_i}{q_j}\right]_p - [Q_i]_p \cdot \left\lfloor \sum_{j_i \leq j < j_{i+1}} \left[ \frac{q_j}{Q_i} \cdot a_j \right]_{q_j} \cdot q_j^{-1} \right\rceil \mod p    
\end{equation}

\subsection{Homomorphic authenticators}
Let $\mathcal{M}$ be a plaintext space, our variation of the construction stated \cite{KLSW2021} is a kind of primitive called homomorphic authenticator (HA), introduced in \cite{Gennaro2013}. A homomorphic authenticator is a tuple of algorithms HA = (KeyGen, Auth, Eval, Ver) which we proceed to describe, adapting the concept to the multigroup setting:

 \begin{enumerate}
     \item KeyGen: on input $p$ and a set $I$, the parties in the set $I$ run the protocol KeyGen, which outputs a joint public key $jpk$, a joint secret key $jsk$ and a joint encrytpion key $jek$. Further, each party $i \in I$ gets shares $pk_i$, $sk_i$ and $ek_i$ of the joint public, secret and encryption keys respectively.
     \item Auth: Given a message $m\in \mathcal{M}$ associated with an identifier $\tau$, it will return an error-detection authenticator $\gamma$.
     \item Eval: It evaluates a deterministic function $f$ on authenticated data $\Gamma = (\gamma_1, \dots, \gamma_n)$ where each $\gamma_i$ authenticates the i-th input $m_i \in \mathcal{M}$.
     \item Ver: It checks that a resulting $m'$ with authentication $\gamma'$ is the correct output of the program $P = (f; \tau_1, \dots, \tau_n)$. 
  \end{enumerate}

A HA must satisfy a series of properties, namely: authentication and evaluation correctness and authenticator security, which are described below:
\begin{enumerate}
    \item Authentication correctness: HA authenticates correctly if, for any message $m\in \mathcal{M}$ with identifier $\tau$, the following equality holds:
    \begin{equation}
        P\left( \text{Ver}(m, \text{Id}_\tau, \gamma; sk) = 1 \,|\, (sk, pk) \leftarrow \text{KeyGen}, \gamma \leftarrow \text{Auth}(m, \tau, sk) \right) = 1        
    \end{equation}
    for $\text{Id}_\tau = (\text{Id}, \tau)$ the labelled program associated with the identity.
    
    \item Evaluation correctness: a HA evaluates correctly if, for any key pair $(sk, pk) \leftarrow \text{KeyGen}$, any fixed function $f$ and any set $\{(P_i, m_i, \gamma_i)\}_{0 \leq i \leq n}$ such that for all $0 \leq i \leq n$, $\text{Ver}(m_i, P_i, \gamma_i, sk) = 1$ let $m^\star = f(m_1, \dots, m_n)$, $P^\star = f(P_0, \dots, P_n)$ and $\gamma^\star = \text{Eval}(f, \gamma_0, \dots, \gamma_n, pk)$, then: 
    \begin{equation}
        \text{Ver}(m^\star, P^\star, \gamma^\star, sk) = 1
    \end{equation}
    
    \item Authenticator security: let us consider the following game $\text{ForgeGame}^A$ between an attacker $A$ and a challenger:
    \begin{enumerate}
        \item The challenger generates keys $sk, pk$ and initializes a list of authenticated inputs $T = \emptyset$.
        \item The attacker $A$ can adaptively submit arbitrarily many authenticator queries $(m, \tau)$ to the challenger. On each query, if there is some $(m, \tau) \in T$, then the challenger aborts. Else, it updates $T = T \cup \{(m, \tau)\}$ and replies $\gamma \leftarrow \text{Auth}(m, \tau)$.
        \item The attacker $A$ outputs a forgery $(m^\star, P^\star = (f^\star, \tau^\star_1, \dots, \tau^\star_n), \gamma^\star)$. The output of the game is $1$ if, and only if, $\text{Ver}(m^\star, P^\star, \gamma^\star) = \text{accept}$ and one of the conditions below holds:
        \begin{enumerate}
            \item Type I forgery: there is some $1 \leq i \leq n$ such that the label $\tau^\star_i$ does not appear in $T$. 
            \item Type II forgery: The set $T$ contains tuples $(m_1, \tau^\star_1), \dots, (m_n, \tau^\star_n)$ such that 
            \begin{equation}
                f^\star(m_1, \dots, m_n) \neq m^\star
            \end{equation}
        \end{enumerate}
    \end{enumerate}
    We say that a HA is secure if, for any probabilistic polynomial-time attacker $A$, we have that
    \begin{equation}
        P(\text{ForgeGame}^A(\text{HA}, \lambda)=1) \leq \text{negl}(\lambda).        
    \end{equation}
\end{enumerate}

\section{Verifiable encodings}
\label{sec:verenc}
Our approach is inspired by the proposal Veritas \cite{Chatel2022} and the use of homomorphic authenticators started by Gennaro and Wichs in \cite{Gennaro2013}. Here we will use a homomorphic authenticator called replication encoding which can be described, informally, as follows: for each input message, one generates an extended vector a random half of which is filled with challenge values whereas the other half is filled with copies of the original message. This extended vector is then encrypted and offloaded to the server. The server will not be able to decide which parts are copies and which parts are challenges, therefore it will impossible for it to tamper with computations without being detected.

Let $\lambda$ be a power-of-$2$ security parameter, $F$ be a variable length pseudorandom function (PRF), and $H$ a collision-resistant hash function (CRHF). Let us also consider a homomorphic encryption scheme HE = (KeyGen Enc, Dec, Eval). The replication encoding authenticator is defined by the following five algorithms:

\begin{enumerate}
    \item KeyGen: given a power-of-$2$ security parameter $\lambda$, this algorithm sets up a PRF and a CRHF to achieve at least $\lambda$-security. It also generates the public and encryption keys.

    \item SetGen: this algorithm generates a challenge set $S \subset \{0, \dots, \lambda\}$ such that $|S| = \lambda / 2$. Since we deal with scenarios involving several users, SetGen will be a multiparty scheme involving each participant in the computations.

    \item Auth: for an input vector $m$ with identifier $\tau$, it proceeds sequentially for each component by initialising an empty vector $M$. For the i-th component $m_i$ of $m = (m_1, \dots, m_k)$, it creates an extended vector $M_i = (m_{i,1}, \dots, m_{i, \lambda})$ where each of its components is either $m_i$ or a challenge value of the form $F(\tau, j)$ if $j$ belongs to the challenge set $S$, to be precise:
    \begin{equation}
         m_{i,j}= 
        \begin{cases}
            m_i & \text{if } j \notin S \\
            F(\tau, j) & \text{otherwise}
        \end{cases}   
    \end{equation}
    The vector $M_i$ is appended to $M$. The algorithm Auth also sets a value $\eta = F(\tau)$, and returns the pair $\gamma = (\text{Enc}(M, pk), \eta) = (ct, \eta)$.

    \item Eval: it takes $n$ inputs, represented by the vector of authentications $\Gamma$ such that $\Gamma_k = (ct_k, \eta_k)$ for $1 \leq k \leq n$. For a function $f$, it computes $ct' = \text{Eval}(f, ct_1, \dots, ct_n, evk)$ by evaluating homomorphically the corresponding circuit on inputs $\{ct_1, …, ct_n\}$. A hash tree $f^H$ is used to ensure that the server has used the appropriate identifiers for all the inputs of the program. It returns the pair $\gamma' = (ct', \eta')$, for $\eta' = f^H(\eta_1, \dots, \eta_n)$.

    \item Ver: this algorithm takes a labelled program $P = (f; \tau_1, \dots, \tau_n)$ and proceeds sequentially:
    \begin{enumerate}
        \item Offline precomputes the challenge values using both the identifier and $F$ and evaluates the function $f$ on them: 
        \begin{enumerate}
            \item $\forall j \in S, \forall k \in \{1, \dots, n\}: r_{k,j}=F(\tau_k, j)$.
            \item $\forall j \in S: \overline{r}_j = f\left(\{r_{k,j}\}_{k \in \{1, \dots, n\}}\right)$
        \end{enumerate}
        \item It computes the value $\eta_k = F(\tau_k)$ for $k \in \{1, \dots, n\}$. Then it computes $\eta^\star = f^H(\eta_1, \dots, \eta_n)$. If $\eta^\star \neq \eta'$, it outputs $0$.
        \item It decrypts the ciphertext $ct'$ to the plaintext vector 
        \begin{equation}
            M^\star = \text{Dec}(ct', sk) = (m_{1,1}, \dots m_{1,\lambda}, \dots, m_{k, \lambda})
        \end{equation}
        It checks for all slots in the challenge set $S$ that the output matches the pre-computed challenge. If not, it outputs $0$: $\forall j\in S$: $m_{1,j} \neq \overline{r}_j \Rightarrow \text{Return }0$.
        \item Finally, it ensures that all other slots evaluate to the same value $m'$. If not, it outputs $0$: $\forall j \in \{0, \dots, \lambda\} - S$: $m_{1,j} \neq m' \Rightarrow \text{Return }0$.
        \item If all the above passes, it outputs $1$.
    \end{enumerate}
\end{enumerate}

The following theorem states that a malicious cloud has a negligible probability of tampering with the results of the computation without being detected:

\begin{theorem}[Theorem IV.1, \cite{Chatel2022}]\label{thmChatel}
Let $\lambda$ be a power-of-$2$ security parameter. If the PRF $F$, the CRHF $H$, and the associated homomorphic scheme are at least $\lambda$-bit secure, then the replication encoding authenticator is a secure authenticator.
\end{theorem}

\section{Verifiable multigroup encryption schemes}
\label{sec:vermghe}
Multigroup homomorphic encryption (MGHE) is a cryptographic scheme that was introduced in the paper \cite{KLSW2021}. It extends the concepts of multiparty and multikey homomorphic encryption and is designed to support several parties. The scheme enables a group of parties to collaboratively generate a public key, which can be used for encryption, thus resembling a multiparty scheme. Furthermore, it allows computations to be performed on encrypted data, even when the input ciphertexts were encrypted under different keys, similar to a multikey scheme. Consequently, multigroup homomorphic encryption offers the benefits of both primitives. In this section, we will introduce a variation allowing subsequent verification.

\subsection{Mutigroup homomorphic encryption schemes}
Let $\mathcal{M}$ be a plaintext space. An multigroup homomorphic encryption scheme over a plaintext space $\mathcal{M}$ is a tuple of algorithms MGHE = (Setup, KeyGen, Enc, Dec, Eval) where:
\begin{enumerate}
    \item Setup: on input the security parameter $\lambda$, this algorithm outputs a public parameter set $pp$.
    \item KeyGen: initially the parties in a set $I$ hold $pp$ and run this protocol. In the end, it outputs a public key $jpk$, and each party $i\in I$ obtains a private share $sk_i$. It also outputs a joint encryption key $jek$ with the associated shares $ek_i$ for $i \in I$.
    \item Enc: given a public encryption key $jek$ and a message $m \in \mathcal{M}$, this algorithm outputs a ciphertext $ct$.
    \item Dec: given a ciphertext $ct$ and the associated secret keys, this algorithm outputs a message $m\in \mathcal{M}$.
    \item Eval: given a function $f:\mathcal{M}^n \to \mathcal{M}$, a collection of ciphertexts $ct_1, \dots, ct_n$, and joint public keys $jpk_1, \dots, jpk_n$, it outputs a ciphertext $ct$.
\end{enumerate}

\subsection{Verifiable  schemes}
We introduce our proposal for a verifiable MGHE scheme (VMGHE), using the replication encoding authenticator. Such kind of scheme is built upon the non-verifiable equivalents of \cite{KLSW2021}, based in the BFV scheme, by including an algorithm SetGen and by replacing the algorithms Enc and Dec with Auth and Ver, respectively. Therefore, a verifiable multigroup homomorphic encryption scheme is a tuple of algorithms VMGHE = \{Setup, KeyGen, SetGen, Auth, Eval, Ver\} which we describe below:

\begin{enumerate}
    \item Setup: for $N > 0$, let us write $R = \mathbb{Z}[x]/(x^N + 1)$ and $R_q = \mathbb{Z}_q[x]/(x^N + 1)$ for $q\in \mathbb{Z}$. Let $\lambda$ be a power-of-$2$ security parameter. Let $p$ be the plaintext space modulus and $q$ the ciphertext space modulus. Write $q^\star = q \cdot q'$ for $q'\in\mathbb{Z}$ such that $\gcd(q, q') = 1$. Let $\chi$ be the key distribution over $R$ and $\sigma > 0$ an error parameter. Let $D(\sigma)$ a distribution over $R$ which samples $N$ coefficients independently from a discrete Gaussian distribution of variance $\sigma^2$. We sample a vector $a \leftarrow U(R_q^{k^\star})$ and take homomorphic gadget decompositions $h: R_q \to R^k$ and $h^\star: R_{q^\star} \to R^{k^\star}$ with gadget vectors $g \in R_q^k$ and $g^\star \in R_{q^\star}^{k^\star}$ respectively. The parameter vector is $pp = (N, q, q', \chi, \sigma, a, h, g, h^\star, g^\star)$.

    \item KeyGen: concerning the encryption key, the protocol KeyGen receives as parameters the vector $pp$. Each party $i$ sets the secret and the encryption keys as follows: 
    \begin{enumerate}
        \item For the secret key, $i$ samples $s_i \leftarrow \chi$ and sets the secret key as $sk_i = s_i$.
        \item For the individual encryption key, $i$ takes $e_{0,i} \leftarrow D^{k^\star}(\sigma)$ computes $b_i = -s_i \cdot a + e_{0,i} \mod q$ and sets the individual encryption key as the pair $ek_i=(b_i[0], a[0])$, using the notation $v[0]$ for the 0-th component of vector $v$.
    \end{enumerate}

    For the public key, the protocol KeyGen receives as parameters a secret key $s_i$. Each party $i$ sets the public key $pk_i = (b_i, v_{0,i}, v_{1,i}, v_{2,i})$ by running the following algorithm:
    \begin{enumerate}
        \item Samples $v_{0,i} \leftarrow U(R^k_q), e_{1,i} \leftarrow D^k(\sigma)$ and lets $v_{1,i} = -s_i \cdot v_{0,i} - r_i \cdot g + e_{1,i} \mod q$.
        \item Samples $r_i \leftarrow \chi, e_{2,i} \leftarrow D^{k^\star}(\sigma)$ and lets $v_{2,i} = -r_i \cdot a + s_i \cdot \lfloor(p/q’) \cdot g^\star\rceil + e_{2,i} \mod q$.
    \end{enumerate}

    Finally, the joint public key for a group $I_l$ is given by the tuple $jpk_l = (\beta_l, \nu_{0,l}, \nu_{1,l}, \nu_{2,l})$, where $\beta_l = \sum_{i \in I_l} b_i$, $\nu_{0,l} = \sum_{i \in I_l} v_{0,i}$, $\nu_{1,l} = \sum_{i \in I_l} v_{1,i}$ and $\nu_{2,l} = \sum_{i \in I_l} v_{2,i}$. The joint encryption key of the group $I_l$ is given by $jek_l = (\beta_l[0], a[0])$. Finally, one considers the individual secret keys $s_i$ as additive shares of an \textit{ideal} group secret key: $jsk_l = \sum_{i\in I_l} sk_i$.
    
    \begin{rmk}
    We note that the public keys are almost linear with respect to the elements $r_i$ and $s_i$, meaning that the joint public key $jpk_l$ satisfies the following condition, where $r=\sum_{i\in I_l}r_i$:
    \begin{equation}
    \begin{cases}
        \beta_l &\approx -jsk_l \cdot a \mod q \\
        \nu_{1,l} &\approx -jsk_l \cdot \nu_{0,l} -r \cdot g \mod q \\
        \nu_{2,l} &\approx - r \cdot a + jsk_l \cdot \lfloor(p/q’) \cdot g^\star\rceil \mod q
    \end{cases}
    \end{equation}
    This condition justifies the fact that combining individual keys yields valid joint keys for the multigroup version of the BFV encryption scheme.  
    \end{rmk}

    \item SetGen: in our proposal for verification in more general settings, it will not be possible for the participants to set their own challenge set since, in the verification stage, different challenge sets may lead to different verification results. Therefore, participants, no matter the group, will be required to set a common challenge set $S$ to be able to verify the results obtained from the server. 

    There exist several ways to define the mechanism, for example using a scheme similar to secret sharing could be a good option. Nevertheless, one will use the fact that public keys for the homomorphic encryption scheme were created right before the generation of the challenge set to use these keys to send encrypted shares of the common challenge set from each user to the rest of the users.

    In SetGen, each participant $i$ will create a share of the set $S_i$, and send it encrypted; then each participant will decrypt their associated shares and join them to generate the set. A description of the mechanism, SetGen$(S_i, \{ek_j\}_{j \neq i})$, follows:

    \begin{enumerate}
        \item Each participant $i$ defines a share $S_i \in \{0, 1\}^{\lambda/2}$. This can be done by coin-flipping.
        \item Each participant $i$ encrypts $S_i$ with all other participant’s encryption keys and broadcasts the vector $\{(\text{Enc}(S_i, ek_j), ek_j)\}_{j \neq i}$. We observe that in this step, users are required to use the MGHE encryption without authentication, that is: given an $R_p$-encoding of $S_i$, each user will compute $\mu_i = t \cdot ek_i + ((q/p) \cdot S_i + e_0, e_1) \mod q$.
        \item Each participant decrypts their corresponding share of the common challenge set $S$ from the list of broadcast vectors. As for encryption, this step does not require any verification, so once a user $i$ receives a message $\mu_i$ it will decrypt it by computing $S_i = \lfloor (p/q) \cdot (c_0 + c_1 \cdot s_i) \rceil \mod p$.
        \item Each participant defines the common challenge set as $S = [\sum_{i} S_i]_2$.
    \end{enumerate}
    
    \item Auth: we follow the replication algorithm described in \hyperlink{section.3}{Section 3}. Given the joint encryption key of group $I$, $jek$, and a message $m\in R_p$, the Auth will produce a pair $\gamma = (ct, \eta)$ as follows: for the extended message $M \in R_p$, generated with the replication encoding, $ct = t \cdot jek + ((q/p) \cdot M + e_0, e_1) \mod q$, for $t \leftarrow \chi$ and $e_0, e_1 \leftarrow D(\sigma)$; $\eta = F(\tau)$, for $\tau$ the identifier of the message $m$.

    Ciphertexts produced by a group of individuals $I_i$ which use the BFV algorithm have the form $ct_i = (c_{0,i}, c_{1, i}) \in R_q \times R_q$. This ciphertext $ct_i$ can be expanded to a multigroup ciphertext involving several groups, $I_1, \dots, I_n$, by doing
    \[
    ct_i = (\underbrace{c_{0,i}, 0, \dots, c_{1,i}}_{i}, 0, \dots, 0) \in R_q^{n+1}
    \]
    Therefore all multigroup ciphertexts will have the structure of a tuple $ct = (c_0, c_1, \dots, c_n)$ which can be decrypted using the ideal secret key $(1, jsk_1, \dots, jsk_n)$ for $jsk_k$ the secret key of group $I_k$, $1 \leq k \leq n$, indeed: $M = \langle ct, (1, jsk_1, \dots, jsk_n) \rangle$.

    \item Eval: this mechanism does not change with respect to the proposal described in \hyperlink{section.3}{Section 3}: it takes a vector of authentications $\Gamma = (\gamma_1, \dots, \gamma_n)$, with $\gamma_j = (ct_j, \eta_j)$ for $0\leq j < n$, and computes $ct'$ by evaluating the circuit associated with the function $f$ on inputs $ct_1, \dots, ct_n$. For the product, we adapt the procedure in \cite{KKLSS2022} for an RNS-friendly product combined with relinearization to the multigroup setting, having in mind that the indices do not describe keys from individual users, but joint keys generated by groups of users:

    On input two ciphertexts $ct = (c_i)_{0\leq i \leq n}$ and $ct' = (c'_i)_{0\leq i \leq n}$ together with a collection of joint public keys $\{jpk_j = (\beta_j, \nu_{0,j}, \nu_{1,j}, \nu_{2,j})\}_{0\leq j\leq n}$ it outputs a ciphertext $ct^\star = (c^\star_i)_{0\leq j \leq n}$:

    \begin{enumerate}
        \item for $0 \leq j \leq n: ct''_j \leftarrow \left\lfloor \frac{q'}{q} c'_j \right\rceil \mod q$
       \item $c^\star_0 \leftarrow \left\lfloor \frac{t}{q'} \cdot c_0c''_0\right\rceil \mod q$
       \item for $1 \leq j \leq n: c^\star_j \leftarrow \left\lfloor \frac{t}{q'} \cdot (c_0c''_j + c_jc''_0)\right\rceil \mod q$
       \item $z \leftarrow \sum_{1\leq i\leq n} h^\star(c_i) \odot \nu_{2,i} \mod q$
        \item $w \leftarrow \sum_{1\leq j\leq n} h^\star(c_j) \odot \beta_j \mod q$
        \item for $1 \leq i,j \leq n$:
        \begin{enumerate}
            \item $c_j^\star \leftarrow c_j^\star + \langle h^\star(c''_j), z \rangle \mod q$
            \item $(c_0^\star, c_i^\star) \leftarrow (c_0^\star + \langle h(\langle h^\star(c_i), w \rangle),  \nu_{1,j}\rangle , c_i^\star + \langle h(\langle h^\star(c_i), w \rangle), \nu_{0,j}\rangle) \mod q $ 
        \end{enumerate}
    \end{enumerate}

    We observe that the output of the above procedure is a ciphertext $ct_{prod} \in R_q^{n+1}$. For the addition of two ciphertexts $ct=(c_i)_{0\leq i \leq n}$ and $ct'=(c'_i)_{0\leq i \leq n}$, the resulting output is the ciphertext $ct_{add} = (c_i + c'_i)_{0\leq i \leq n} \mod q$.

    \item Ver: the verification process is composed of two steps: an offline phase where each participant precomputes the challenge values according to function $f$, following algorithms in \hyperlink{section.3}{Section 3}, and an online stage where a distributed decryption mechanism is used. 
   \newpage
    For the distributed decryption, one has $ct = (c_0, \dots, c_n)$ a multigroup ciphertext corresponding to an ordered set of groups $(I_1, \dots, I_n)$ and an error parameter $\sigma’ > 0$, now:
    \begin{enumerate}
        \item Each participant $i \in I_j$ samples $e_i’ \leftarrow D(\sigma’)$ and broadcasts $\mu_{j,i} = c_j \cdot s_i + e’_i \mod q$.
        \item Each participant $i \in I_j$ computes and broadcasts $\mu_j = \sum_{i \in I_j} \mu_{j,i} \mod q$.
        \item Merging yields $\mu = c_0 + \sum_{1 \leq j \leq n} \mu_j \mod q$.
        \item Finally, each participant decrypts $m = \lfloor (p/q) \cdot \mu\rceil \mod p$.
    \end{enumerate}

    Once the message has been decrypted, every participant can run the verification algorithm described in \hyperlink{section.3}{Section 3}.
\end{enumerate}

\begin{rmk}
We observe that Auth is the non-verifiable equivalent algorithm Enc including the replication authenticator. Similarly, the Ver algorithm corresponds to the non-verifiable Dec algorithm, including the verification procedure of the replication authenticator.
\end{rmk}

\section{Security and correctness}

\subsection{Security}
Let $I_1, I_2, \dots, I_n$ be sets of parties and let $\mathcal{I} = \bigcup_{1\leq j \leq n}I_j$. Let $\mathcal{A} \subset \mathcal{I}$ be a set of attackers and $\mathcal{H} = \mathcal{I} - \mathcal{A}$. An MGHE scheme is semantically secure if the advantage of $\mathcal{A}$ in the following game is negligible for any probabilistic polynomial-time adversary $\mathcal{A}$: 
\begin{enumerate}
    \item Setup: The challenger generates a public parameter $pp \leftarrow \text{Setup}(\lambda)$.
    \item Key Generation: Adversary plays with an honest challenger the key generation protocols KeyGen for all $1 \leq j \leq n$. At the end of the protocol, the adversary gets the secret shares 
    \[
    \{sk^i_j: i \in \mathcal{A},\, 1\leq j \leq n\}
    \]
    and the challenger receives $\{sk^i_j: i \in \mathcal{H},\, 1\leq j \leq n\}$.
    \item Challenge: The adversary chooses messages $m_0, \dots, m_1 \in \mathcal{M}$ and an index $j$ such that $I_j \notin \mathcal{A}$ and sends them to the challenger. The challenger samples a random bit $b \in \{0, 1\}$ and sends $\text{Enc}(m_b, pk_j)$ back to the adversary.
    \item Output: The adversary $\mathcal{A}$ outputs a bit $b'\in\{0, 1\}$. We define the advantage of the adversary as $\left| P(b = b') - \frac{1}{1} \right|$
\end{enumerate}

The MGHE scheme introduced by \cite{KLSW2021} is semantically secure under the RLWE assumption with parameter set $(N, q, \chi, \sigma)$. The proof of this statement follows the reasoning that the reader will find in \cite{KLSW2021}. 

Therefore the security of this VMGHE scheme follows from the semantic security of the non-verifiable scheme combined with Theorem \ref{thmChatel}.

\subsection{Correctness}

Let $pp \leftarrow \text{Setup}(\lambda)$, let $jpk_1, \dots, jpk_n$ public keys generated by sets $I_1, \dots, I_n$ and $jsk_1, \dots, jsk_n$ and $jek_1, \dots, jek_n$ the corresponding joint secret and encryption keys. For any $m_1,\dots m_n \in \mathcal{M}$ and indices $1 \leq i \leq n$ let $ct_i \leftarrow \text{Enc}(m_i, jek_i)$. For any function $f: \mathcal{M}^n \to \mathcal{M}$. Then an MGHE scheme is correct if the following holds:
\begin{equation}
    \text{Dec}(jsk_1, \dots jsk_n; \text{Eval}(jpk_1, \dots, jpk_n; f; ct_1, \dots, ct_n)) = f(m_1, \dots, m_n)    
\end{equation}
with overwhelming probability in $\lambda$.

Proving the correctness of this algorithm is reduced to proving the correctness of the product and relinearization algorithm, for which we refer the reader to the analysis made in \cite{KKLSS2022} which can be trivially extended to the multigroup setting. The correctness of the authenticator, both when it comes to authentication and evaluation, follows from the correctness of the underlying encryption scheme.

\section{Working without a CRS}

Multiparty, multikey and multigroup homomorphic cryptography has become increasingly popular for achieving secure computation with reduced communication costs. All these schemes rely on the common random string (CRS) model that requires a trusted party to distribute the CRS to each participant. This model is considered an ideal form of homomorphic encryption, but it could weaken the semantic security of this kind of scheme.

The CRS links all the parties' ciphertexts under different keys to facilitate correct computation. In an effort to eliminate the CRS's role, Kim, Lee, and Park proposed an alternative scheme in \cite{KLP2018}. In their scheme, parties share their public keys after the key generation step to establish relations among their keys, akin to the threshold fully homomorphic encryption scheme \cite{Asharov2012}. Notably, each user generates their keys independently.

In contrast, prior schemes involve each user generating their key pair, encrypting their message using single-key encryption, and publishing them together at once without previous interaction. However, this approach is designed for a fixed number of users, which may limit its dynamic properties, such as the ability for users to join or leave the computation freely.

The construction of a verifiable MGHE scheme without a CRS follows the steps described in \hyperlink{section.4}{Section 4}, therefore we will focus on the key generation and the product and relinearization operation and refer to \hyperlink{section.4}{Section 4} for authentication and verification.

In order to remove the CRS, each party $i\in I$ generates its keys following a variation of the algorithm described in \hyperlink{section.4}{Section 4}:
\begin{enumerate}
    \item For the individual secret key, $i\in I$ samples $s_i \leftarrow \chi$ and defines the secret key as $sk_i = s_i$.
    \item For the individual encryption key, the party $i\in I$ samples $a_i \leftarrow U(R^k_q)$ together with $e_{0,i} \leftarrow D^k(\sigma)$ and computes $b_i = -s_i \cdot a_i + e_{0,i} \mod q$. Then $ek_i = (b_i[0], a_i[0])$.   
\end{enumerate}

For the public and joint keys, each party $i\in I$ runs variations of the schemes in \hyperlink{section.4}{Section 4}. Indeed, individual public keys are generated as follows: each party $i\in I$ sets the public key $pk_i = (b_i, v_{0,i}, v_{1,i}, v_{2,i})$ by running the following algorithm:
\begin{enumerate}
    \item Samples $v_{0,i} \leftarrow U(R^k_q), e_{1,i} \leftarrow D^k(\sigma)$ and lets $v_{1,i} = -s_i \cdot v_{0,i} - r_i \cdot g + e_{1,i} \mod q$.
    \item Samples $a_i \leftarrow U(R^k_q), r_i \leftarrow \chi, e_{2,i} \leftarrow D^k(\sigma)$ and lets $v_{2,i} = -r_i \cdot a_i + s_i \cdot \lfloor(p/q’) \cdot g^\star\rceil + e_{2,i} \mod q$.
\end{enumerate}

The joint public key for a group $I_l$ is given by the tuple $jpk_l = (\beta_l, \nu_{0,l}, \nu_{1,l}, \nu_{2,l})$, where $\beta_l = \sum_{i \in I_l} b_i$, $\nu_{0} = \sum_{i \in I_l} v_{1,i}$, $\nu_{1} = \sum_{i \in I_l} v_{0,i}$ and $\nu_{2} = \sum_{i \in I_k} v_{2,i}$. The joint encryption key of the group $I_l$ is given by $jek_l = (\beta_l[0], \alpha_l[0])$, for $\alpha_l = \sum_{i \in I_l} a_i$, which behaves like an Irwin-Hall distribution. All groups broadcast their joint encryption keys.

For each group $I$, each party $i \in I$ generates public keys $pk_i^j$, for $1\leq j \leq n$, using all other groups joint encryption keys $jek_j = (\beta_j[0], \alpha_j[0])$ running the following algorithm:
\begin{enumerate}
    \item Sample $v^j_{0,i} \leftarrow U(R^k_q)$, $e_{1,i} \leftarrow D^k(\sigma)$ and let $v^j_{1,i} = -s_i \cdot v^j_{0,i} - r_i \cdot g + e_{1,i} \mod q$.
    \item Sample $r_i \leftarrow \chi$, $e_{2,i} \leftarrow D^k(\sigma)$ and let $v^j_{2,i} = - r_i \cdot \alpha_j + s_i \cdot \lfloor(p/q’) \cdot g^\star\rceil + e_{2,i} \mod q$.
\end{enumerate}

Then each group $I_l$ generates a joint public key $jpk_l^j$ for all other groups by setting $jpk_l^j = (\nu^j_{0,l}, \nu^j_{1,l}, \nu^j_{2,l})$, where $\nu^j_{0,l} = \sum_{i \in I_l} v^j_{0,i}$, $\nu^j_{1,l} = \sum_{i \in I_l} v^j_{1,i}$ and $\nu^j_{2,l} = \sum_{i \in I_l} v^j_{2,i}$.

Concerning the product and relinearization operation, on input two ciphertexts $ct = (c_i)_{0\leq i \leq n}$ and $ct' = (c'_i)_{0\leq i \leq n}$ together with a collection of joint public keys $\{\beta_i, jpk^j_i = (\nu^j_{0,i}, \nu^j_{1,i}, \nu^j_{2,i})\}_{0\leq i, j\leq n}$:

\begin{enumerate}
    \item for $0 \leq j \leq n: c''_j \leftarrow \left\lfloor \frac{q'}{q} c'_j \right\rceil \mod q$.
    \item $c^\star_0 \leftarrow \left\lfloor \frac{t}{q'} \cdot c_0c''_0\right\rceil \mod q$.
    \item for $1 \leq j \leq n: c^\star_j \leftarrow \left\lfloor \frac{t}{q'} \cdot (c_0c''_j + c_jc''_0)\right\rceil \mod q$
    \item $z \leftarrow \sum_{1\leq i\leq n} h^\star(c_i) \odot \nu^j_{2,i} \mod q$; 
    \item $w \leftarrow \sum_{1\leq j\leq n} h^\star(c_j) \odot \beta_j \mod q$
    \item for $1 \leq i,j \leq n$:
    \begin{enumerate}
        \item $c_j^\star \leftarrow c_j^\star + \langle h^\star(c''_j), z \rangle \mod q$
        \item $(c_0^\star, c_i^\star) \leftarrow (c_0^\star + \langle h(\langle h^\star(c_i), w \rangle),  \nu^j_{1,i}\rangle , c_i^\star + \langle h(\langle h^\star(c_i), w \rangle), \nu^j_{0,i}\rangle) \mod q $ 
    \end{enumerate}
\end{enumerate}

The correctness of this version of the product algorithm follows the reasoning of \cite{KKLSS2022} which can be trivially extended to this multigroup setting without CRS.

\section{Future research: hybrid environments}

Fully homomorphic encryption (FHE) is possible, allowing for the evaluation of any function in a homomorphic manner. However, when considering its application in outsourcing computations, the efficiency of the process is debatable. This is due to the fact that the error in a ciphertext depends greatly on the function being evaluated and its expression, as shown by error-growth studies. Furthermore, the handling of noise presents a major challenge in achieving an efficient FHE framework, resulting in common bottlenecks associated with FHE.

Homomorphic encryption is difficult to implement in resource-constrained environments, such as IoT, due to its computational complexity and the associated ciphertext expansion, which significantly increases communication overhead. The limited hardware capabilities of IoT devices, as well as the bandwidth constraints of current IoT communication standards, exacerbate these issues. Hybrid protocols combining homomorphic encryption and symmetric key encryption have been developed to address these concerns.

The three main concerns with efficient fully homomorphic frameworks that have emerged from the development of main FHE generations are the size of the ciphertexts, the complexity of bootstrapping, and the choice of when to bootstrap. Improving these bottlenecks would enhance the efficiency of outsourcing computation. Nonetheless, some of these issues can be addressed by utilizing a hybrid framework.

In a hybrid homomorphic encryption system, data is first encrypted using a symmetric key encryption scheme with a randomly generated key by the IoT device. The data is then encrypted with a homomorphic encryption scheme using the data collector's public key. A symmetric key encryption scheme is less complex and not affected by ciphertext expansion. The intermediate fog nodes can then homomorphically evaluate the decryption circuit of the symmetric key encryption scheme, converting secret key-encrypted data into homomorphically encrypted data that can be processed and sent to the data collector.

We follow \cite{Meaux2017} to introduce the stages which compose a hybrid environment. Let us consider a homomorphic scheme (Hom.KeyGen, Hom.Enc, Hom.Dec, Hom.Eval) and a symmetric scheme (Sym.KeyGen, Sym.Enc, Sym.Dec), then the hybrid environment is composed by the following stages:
\begin{enumerate}
    \item Initialization: using the security parameter $\lambda$, a user $A$ runs the key generation homomorphic KeyGen and the symmetric KeyGen obtaining keys $(pk^H, sk^H)$ and $sk^S$ respectively. User $A$ computes the homomorphic encryption $\text{Hom.Enc}(sk^S, pk^H)$.
    \item Storage: user $A$ computes $\text{Sym.Enc}(m_i)$ and sends it to $B$ together with $\text{Hom.Enc}(sk^S, pk^H)$.
    \item Evaluation: user $B$ homomorphically evaluates the decryption $\text{Hom.Enc}(m_i)$ of $\text{Sym.Enc}(m_i)$. The Eval algorithm is defined only over homomorphic ciphertexts, so $B$ computes 
    \begin{equation}
        \text{Hom.Enc}(\text{Sym.Enc}(m_i)) = \text{Hom.Enc}(\text{Sym.Enc}(m_i), pk^H)    
    \end{equation}
    and then evaluates homomorphically the symmetric decryption 
    \begin{equation}
        \text{Hom.Enc}(m_i) = \text{Hom.Eval}(\text{Sym.Dec}, \text{Hom.Enc}(\text{Sym.Enc}(m_i)), \text{Hom.Enc}(sk^S), pk^H).        
    \end{equation}
    \item Computation: user $B$ homomorphically executes $f$ on $A$’s encrypted data.
    \item Results: user $B$ sends a compressed encrypted result of the data treatment $\text{Hom.Enc}(f(m_i))$ and $A$ just needs to decrypt it.
\end{enumerate}

The single-user situation can be adapted to a multigroup setting by modifying the first stage as follows: in the initialization stage, all users are required to generate the challenge set $S$ for subsequent verification of the results. Furthermore, each group will be required to generate a joint symmetric key together with the keys involved in VMGHE. The joint symmetric key can be generated using a multiparty protocol similar to the algorithm needed to create the challenge set: each user $i$ a group $I$ will generate a random share of the symmetric key which will be sent encrypted, using the VMGHE keys to all other users of $I$. In the end, each user will create a common symmetric key by adding the shares.

The next stages of the hybrid environment can be adapted straightforwardly to the multigroup setting by considering not users $A$ but groups of users $I_j$ for $1 \leq j \leq n$. Here encryptions would be done using the joint encryption keys of VGMHE and each group $I$ would be required to send not only the symmetric encryption of their extended message $M_I$ but also the homomorphic encryption of the joint public key. 

The next stages, namely evaluation, computation and results, follow the steps described above and we only need to add a verification stage, which is done by each user according to \hyperlink{section.4}{Section 4}.

\newpage
\section{Conclusion}
As the main contributions in this paper we have adopted the recent construction MGHE from \cite{KLSW2021} to generate a verifiable variation built upon the ideas from \cite{Chatel2022}. We also adapted the work from \cite{KKLSS2022} to improve the efficiency of the most expensive operations, namely the product and the relinearization algorithms. Furthermore, we have constructed a version that avoids requiring the common reference setting, bypassing a possible decrease in terms of security.

We want to finish this research by pointing out that MGHE is such a plastic construction that one can derive from its equivalent constructions for verifiable MKHE and MPHE by considering the number of parties and groups involved in the computations. Indeed, if we consider that there is only one group of users, the construction introduced in this research becomes a verifiable MPHE; on the other hand, if consider the existence of several groups of users composed of one participant in each case, then what one gets is a verifiable MKHE scheme.

It remains for future work the implementation of the algorithms and ideas presented in this paper, which will bring to the table a benchmarking supporting the affirmation that these algorithms attain a high level of efficiency and time execution.

\section*{Acknowledgements}
The author would like to thank Shreemoy Mishra, from the Research and Innovation Unit at IOV Labs, for the patience shown while reading the article, and for his insightful remarks and comments which improved the quality of the research contained in this paper.


\end{document}